# Simulating a questionnaire on the framing effects in decision-making processes in a serious game
### Research report

Sara Knežević, Mlađan Jovanović

# Contents

# Introduction

The rapid development of technology has introduced new formats of human-computer interaction. Along with the machines themselves, there has been a lot of work on the frontier of input and output devices, which have in turn produced many new forms of media and a whole new field of interactive multimedia.

One of the major mediums that has grown in popularity since its early development is video games. For a long time, video games have been developed and distributed for the purpose of entertainment, however, in the late 2010s, researchers have taken an interest in the characteristics of games and how they can be used for different purposes. The competitive and rule-based aspects of video games, as well as the explorative and problem-solving features make up an environment which incorporates the player's attention, dedication and effort in a virtual world in which the video game is set. Applying these game design elements into non-game contexts is called *gamification*. (Deterding, Dixon, Khaled, & Nacke, 2011)

An important distinction should be made between *gamification* and *serious games*. While gamification includes game design elements, serious games represent an entire game as a whole which is used for non-entertainment purposes. Instead, it leverages the medium to explore certain aspects of interaction, the human psyche, or mechanical skills. Video games are uniquely positioned for this – and increasingly more so as graphics technology keeps improving – since they allow a tight loop of action-reaction. This type of rapid feedback provides fertile ground for many types of experiments which would be impossible or prohibitively difficult to perform in the physical world, and as such serve as strong virtual alternatives.

One of the main features of video games is interactability. For example, by making a strategic decision during a game, a player can determine the course of action of the game. One of the aspects that affect the interactability of video games is the concept of immersion. A video game that is able to produce an immersive experience for the player in which the player believes that they are "actually there" in the game, and that the game is an extension of reality provides an alternate way to explore human behaviours and decision-making processes.

Psychological questionnaires are research instruments consisting of a series of questions for the purpose of gathering information from respondents. They are an effective means of measuring different traits, attitudes, opinions or intentions from different people, usually in the form of written questions or conversation. The term "questionnaire" can be used in two distinct senses: a questionnaire as a research instrument; and a questionnaire as a method of surveying, or a way of collecting information, such as a survey. (McLeod, 2018)

This thesis focuses on bringing a psychological questionnaire on the prospect theory to another level. Prospect theory questionnaires explore decision-making in hypothetical situations. In most cases, these experiments are done in controlled environments and rely on the respondent's imagination to reproduce the situation which is presented to them. With this in mind, the questionnaire lacks the "real life" component, features that would directly affect the decision-making process in one way or another. An increased sense of existing in a life-like environment is thought to magnify the way in which the respondent makes decisions in the situations presented to them. This is described by Cummings and Bailenson as a two-dimensional construct, comprised of a sense of self-location and perceived possibilities to act and it is called *presence*. (Cummings & Bailenson, 2016)

Developing a serious video game which explores hypothetical situations removes the barrier between the respondent and the hypothetical part of the problems at hand by introducing them to the fictional world directly and allowing them to interact with it while seeing the consequences of their decisions instantly. Removing this barrier creates a presence which otherwise wouldn't be as dominant in a situation where the respondent were answering questions on paper or in conversation. Virtual worlds such as ones constructed in video games are the perfect candidate for this – the removal of boundaries by immersion happens naturally to the player, provided that the game is designed well enough to do so.

This serious game introduces an adventurer-like character whose goal is to interact with non-playable characters in a village and solve a quest. All of the questions have the same theme, but present different types of risky situations which affect the player's character directly and the thesis explores how people make decisions when there is an increased element of presence.

# 1  Unity technology for game development

Unity is a game engine and an integrated development environment (IDE) introduced by Unity Technologies in 2005 which is used for creating interactive media (video games, real-time 3D animations, architectural visualizations, etc.). The editor runs on Windows OS and Mac OS X platforms, but has the ability of developing applications for multiple platforms provided the corresponding software development kit (SDK).

## 1.1  Scripting

Unity offers developers two ways in which they program their code: user interface and C#. The scripts are compiled by Mono – an open-source version of Microsoft's .NET framework. This allows the code to be compatible on different platforms. The scripts themselves contain the gameplay, defining the inputs, events and content features of the game. The user interface offers a visual way to interacting with and modifying the code by changing the values of public variables through the Unity editor. This makes it so that developers and game designers have easier access to tweak and modify specific variables without the need of knowing how to write C# code.

C# is an object-oriented and component-oriented programming language that runs in the .NET framework. All C# types, including the primitive types such as *int* and *double*, inherit from a single root *object* type. All types share a set of common operations which makes it so that any type can be stored, transported and operated on consistently. C# supports both user-defined reference types and value types and allows dynamic allocation of objects and in-line storage of lightweight structures. Additionally, it supports generic methods and types, which increase type safety and performance.

.NET is a virtual execution system and a set of class libraries. Source code written in C# is compiled into an intermediate language (IL) which stores its code and resources into an assembly, commonly with the extension *.dll*. IL code produced by the C# compiler conforms to the Common Type Specification (CTS). C# generated IL code interacts with code that was generated from the .NET versions of any CTS-compliant languages. A single assembly may contain multiple modules written in different .NET languages, and the types can reference each other as if they were written in the same language. (Wagner, 2021)

```
using UnityEngine;
using System.Collections;

public class Example : MonoBehaviour {
    void Update() {
        // Move the object forward along its z axis 1 unit/sec
        transform.Translate(Vector3.forward * Time.deltaTime);

        // Move the object upward in world space 1 unit/sec
        transform.Translate(Vector3.up * Time.deltaTime, Space.World);
    }
}
```

*Image 1 Unity code example written in C#*

## 1.2 Rendering

3D rendering is the process of producing an image based on three-dimensional data stored on your computer. The graphics processing unit (GPU) converts 3D wireframe models into 2D images with 3D photorealistic effects. There are two major types of rendering in 3D which are differentiated by their calculation and processing speed: pre-rendered and real-time rendering. Most common in video games is real-time rendering that creates an environment in which scenes occur in real time as players interact with the game.

Unity builds its graphical elements on top of the low-level DirectX library, using the OpenGL library for Windows OS, Mac and Linux and OPenG1 ES and WebGL for mobile and web development.

The DirectX rendering pipeline begins with the input assembler reading the vertices and indices of 3D objects from memory and approximating them by triangle meshes. Each object, along with its size and rotation, is placed in a local coordinate system which is later brought together with the global coordination system using geometric transformations. The virtual camera is then translated to the origin of the world space, which allows *culling* of polygons which are not seen by the camera. Lighting is necessary in order to create a photorealistic scene. Light sources are game objects placed in the world space with specific properties (color, intensity, direction, position and focus). Culling is implemented to decide which objects should be ignored by the renderer according to the *view frustum*. All objects inside the frustum is kept in the scene, while the object that is between the inside and outside of the frustum is partially culled. Finally, the scene is rendered based on perspective projection of the 3D vertices onto a 2D projection window inside the frustum.

Unity has additional rendering steps that occur along with the main DirectX pipeline. *Forward rendering* separates objects which are not affected by lights, opaque objects and transparent objects that are combined to the opaque objects' colors. *Deferred shading* manages geometry buffers to compute geometry before applying lighting. *Pre-pass rendering* restricts the usage of material shaders in deferred shading. *Vertex lit rendering* calculates the lighting on all vertices of objects from all light source objects. (Messaoudi, Simon, & Ksentini, 2015)

## 1.3 Assets

Game assets are representations of any item that can be used within a game or a project. The assets can be imported (such as 3D models, audio files, images), created in a 3rd party software (such as 3DS Max, Blender, Zbrush) or created within Unity (such as animation controllers, render textures, materials).

The asset workflow consists of five steps:

1. Importing – bringing the source files into the Unity editor to work with
2. Creating – placing assets into scenes or game objects and adding scripts to them
3. Building – exporting the completed project into a binary file which can be run
4. Distributing – creating access to the project
5. Loading – users loading and/or using the project along with the assets built within it

Two of the major unity asset types are 3D and 2D assets. 3D assets include characters, animations, vegetation, property items. Unity supports a number of standard model file formats, however its built-in importing chain uses the *.fbx* file format. Other formats, including *.dae*, *.dfx* and *.obj*, are internally converted into the built-in format. Unity's animation tools allow characters to be compatible with animations from different sources.

Each 3D model has its appropriate 3D mesh, which is the main graphics primitive of Unity which are rendered by the engine's components into regular or skinned meshes, trails or 3D lines. While meshes define the geometry of an object, textures are required to supply the details of an object. Textures are bitmap images that are applied over the mesh surface using materials. Materials are created within the Unity editor and use specialized graphics programs called *shaders* which render the texture on the game object. Materials can be applied to meshes, particle systems or GUI textures. Unity imports textures from the most common image formats.

2D assets include sprites, textures, environments, fonts, materials, UI elements and textures. Sprites use textures applied to flat surface meshes which approximate the object's shape. Graphic user interfaces (GUI) are used to display information and interactions that are not directly in the game scene itself and they use standard textures as well. Terrain heightmaps are greyscale textures which describe the height values of the terrain in game. Each pixel has a value ranging from 0 to 1 that corresponds to the shade of grey at that point in the image and represents the height of the terrain.

## 1.4 Unity editor

The Unity editor includes five key views used in development:

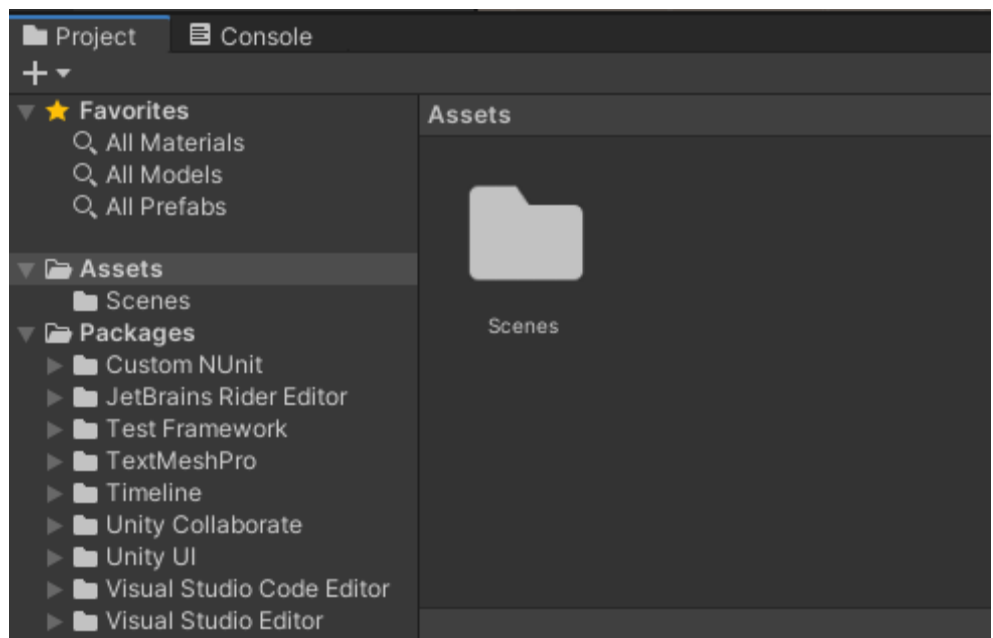

*Image 2 Unity project view*

1. Project view

Project view allows access to all of the necessary files and assets for the game development. It's a file explorer which makes files and folders accessible to be created, modified and deleted. Image 2 shows the project view of a newly created Unity project.

2. Scene view

Scene view shows the game level and is used to place and build the level itself. Image 3 shows a 3D scene with a plane game object and a cube placed on top.

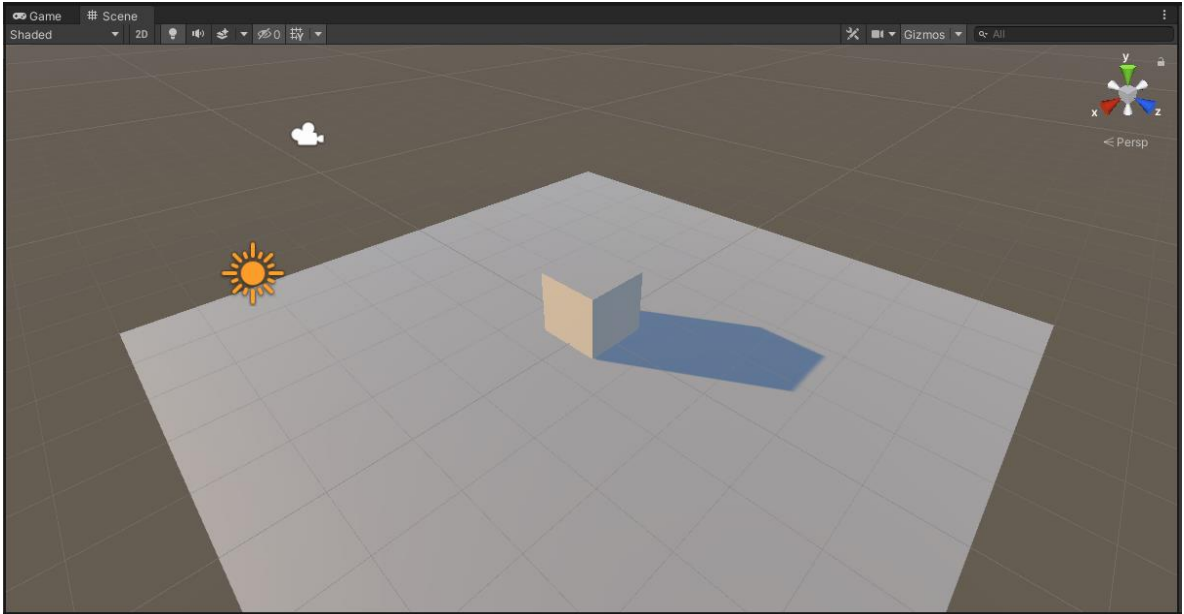

*Image 3 Unity scene view*

3. Game view

Game view shows how the camera renders the created scene based on its location and rotation in the world space. Image 4 shows the camera view of the created game objects in Image 3, as well as additional resolution options in the upper left corner.

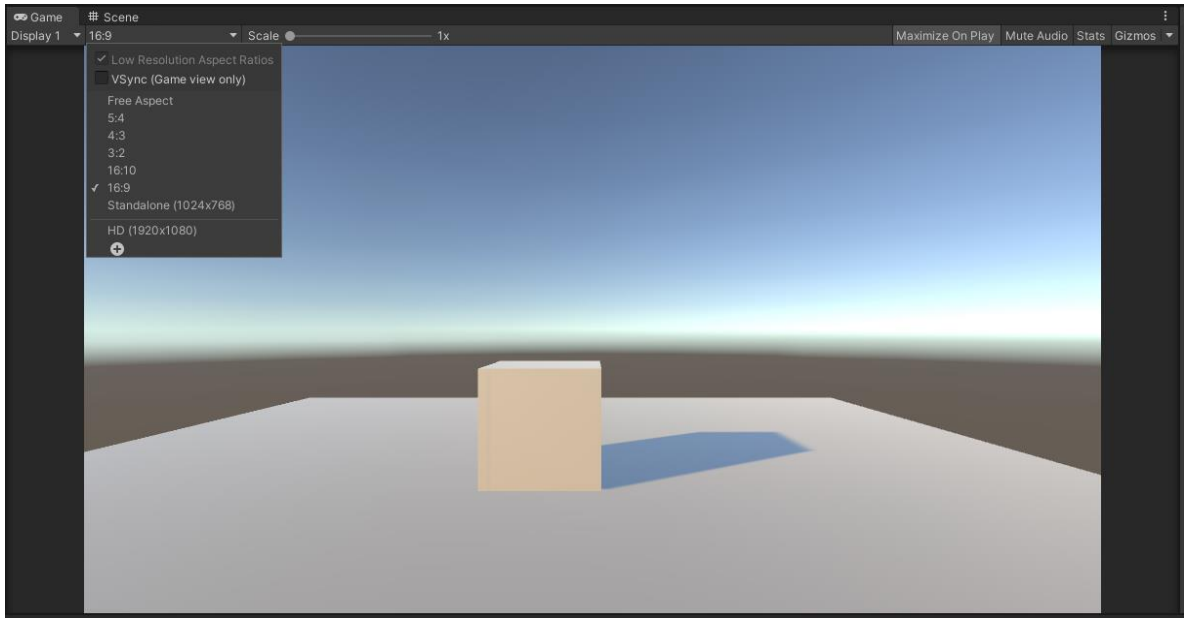

*Image 4 Unity game view*

4. Hierarchy view

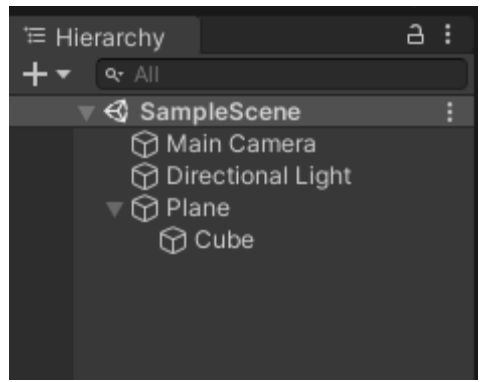

*Image 5 Unity hierarchy view*

Hierarchy view displays all of the game objects that exist in the current scene. It allows creation, accessibility, manipulation and grouping of the objects in order to create the game. Image 5 shows the hierarchy of game objects created for the scene in Image 2.

5. Inspector view

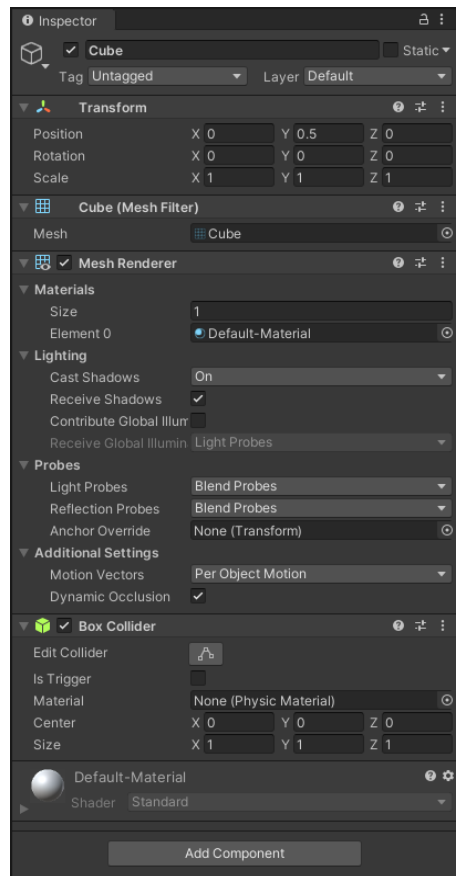

*Image 6 Unity inspector view*

Inspector view displays all of the components and transform values of game objects. The transform component contains information on the object's position, rotation and scale. Some of the components include physics properties, textures, colliders, meshes, etc.

# 2 Research question

The goal of this thesis is to explore how people make decisions in risky situations which are simulated in a serious game. Players are placed in a mid-century fictional world through which they explore and interact with non-playable characters (NPC). The NPCs' role is to present the player with a hypothetical situation in which they have to pick between two different options of action, each of which affect the player's character directly.

Prospect theory represents a form of rational decision-making and introduces an emphasis on the positive and negative outcomes of certain and uncertain (risky) options based on their estimation of the perceived likelihood of each of them. People make decisions based on the presented gains instead of the losses. This is known as *loss-aversion* and it presumes that if two options are presented which have the same value, where one option is presented in terms of gains and the other in terms of loss, the former is more likely to be chosen.

The theory was introduced in 1979 by Daniel Kahneman and Amos Tversky and it is a more psychologically based theory on how people make decisions when compared to the expected utility theory (EUT).

## 2.1 Expected Utility Theory

Expected utility theory (EUT) states that people choose between risky options by comparing the weighted sum of the utility values of payoffs multiplied by their probabilities:

$$U(p) = \sum(x_k)p_k,$$

where U represents the utility of each payoff, p the probability, x the payoff. The utility of an option is the equivalent to its expected utility. EUT is based on three primary principles: (1) overall utility of an option (U) is the expected utility of its outcomes, (2) an option with an additional asset is acceptable if the utility of the integrated option and asset is greater than the utility of that asset alone and (3) the utility is concave, which makes risk aversion equivalent to the concavity of the utility function. (Elgar, 1998)

EUT varies based on the situation. This means that the utility function can be convex (risk-averse) or concave (risk-prone), or be linear and show an indifference towards the different options of action. This theory is based on a number of axioms which allowed for mathematical modelling of decision-making based on the respondent's behavior (descriptive approach) and analysis of each of the principles (normative). One of the axioms implies that the choice between two options should be decided based on the difference between the choices themselves, not external factors which affect both choices.

This theory represents one of the main normative models, but is not as reliable as a descriptive perspective and has since been put into question by multiple economists, psychologists and mathematicians, including Kahneman and Tversky.

Maurice Allais was one of the first people to push the boundaries of expected utility theory. *Allais paradox* is a problem designed to show the inconsistency of actual observed choices with the predictions of EUT. This paradox shows that people are risk-averse to gaining nothing. This phenomenon is known as the zero effect and is a similar concept to the certainty effect, described by Kahneman and Tversky, which suggests that people are attracted to certain and secure options.

Allais paradox describes 4 different lottery options, where the only lottery without a zero outcome (option 1A) is the certain lottery. The decision maker is presented with two options between which they must choose. (Allais, 1953)

Gamble A:

Option 1A: 100% chance of winning $100 million
Option 1B: 89% chance of winning $100 million
      10% chance of winning $500 million
      1% chance of winning nothing

Gamble B:

Option 2A: 89% chance of winning nothing
      11% chance of winning $100 million
Option 2B: 90% chance of winning nothing
      10% chance of winning $500 million

The paradox shows that most people pick option 1A and 2B. EUT implies that in that case, option 1B would be picked over 2B. The expected utilities of the preferred options are greater than the expected utilities of the second options which contradicts the first gamble, where people pick the certain option.

$$Gamble\ A\ expected\ utility:$$
$$1U(\$100M) > 0.89U(\$100M) + 0.1U(\$500M) + 0.01(\$0M)$$

$$Gamble\ 2\ expected\ utility:$$
$$0.89U(\$0M) + 0.11U(\$100M) < 0.9U(\$0M) + 0.1U(\$500M)$$
$$rewritten\ as$$
$$1U(\$100M) < 0.89U(\$100M) + 0.01U(\$0M) + 0.1U(\$500M)$$

A lot of issue with this normative theory has led to the development of descriptive theories which would predict decision-making in risky situations. While normative theories assume that the decision-maker is a rational agent who calculates utility values of each option, descriptive theories rely on human behavior and take into account the probabilities but also lack of information, subjectivity, language misunderstandings, etc. (Damnjanović, 2014)

## 2.2 Prospect theory

Prospect theory is one of the leading descriptive theories on risky decision-making. The two Israeli psychologists introduced the theory (Kahneman & Tversky, Prospect Theory: An analysis of Decision under Risk, 1979) as a more psychologically-based theory compared to expected utility theory. The paper showed that in a laboratory setting, people commonly violate ETU's principles and predictions and proposed a new theory which would better predict how people behave in hypothetical situations and make decisions. Prospect theory itself shows us that people are not perfectly rational agents who make normative decisions and instead have cognitive biases. In other words, our choices are results of our subjective perception of the probability and value of each option.

The probabilities on which the decision-maker decides are not numerical objective values from 0 to 1, but rather subjective probabilities whose value varies as well as the probability. Subjective probabilities don't always correspond to the objective ones. (Gvozdenović & Damnjanović, 2016) The value function differentiates prospect theory from ETU by marking the relationship between the decision-maker's deviation from the reference point. If the value is greater than the reference point, the option is considered a gain, while if it's less than the reference point, it's considered to be a loss. (Kahneman & Tversky, Prospect Theory: An analysis of Decision under Risk, 1979) (Damnjanović, 2014)

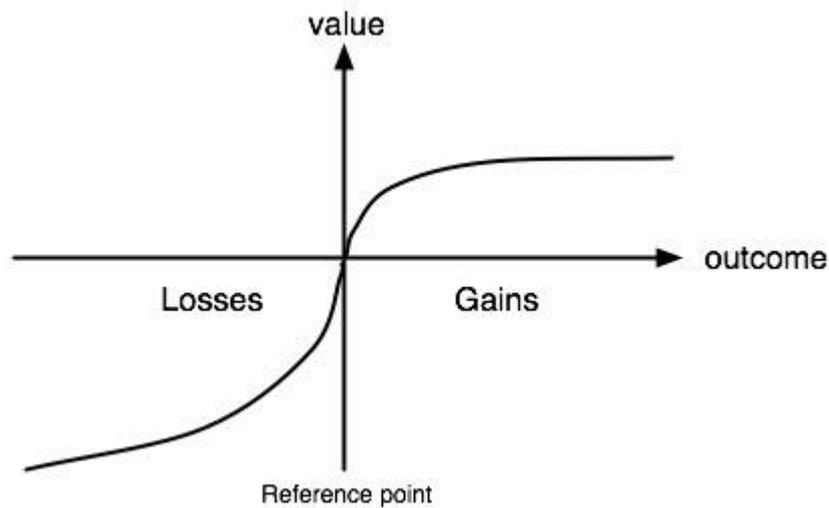

*Image 7 Function of value*

The following effects discovered through laboratory experiments defined in the paper appear to invalidate ETU as a descriptive model and introduce the newly established descriptive theory. (Kahneman & Tversky, Prospect Theory: An analysis of Decision under Risk, 1979)

### 2.2.1 Certainty effect

The certainty effect describes a phenomenon where people overweight results which are considered certain, relative to the outcomes which are probable. An example of this is the Allais paradox's first option described above, where people mostly pick the first certain option, rather than the risky one.

### 2.2.2 Reflection effect

The preferences between negative prospects mirror the preferences between the positive prospects, meaning that risk aversion in the positive domain means risk seeking in the negative. In the positive domain, the certainty effect contributes to risk averse preferences which have a certain result, over other probabilistic options, while in the negative domain the same effect leads to risk seeking preferences over smaller losses which are certain.

### 2.2.3 Isolation effect

People often ignore components which are shared between options, marking them as irrelevant and focusing only on the components which are unique to each option. This causes a problem because those components can be decomposed in various ways and thus lead to different results of each option. By introducing a dependency between the options without changing the results or probabilities a new certainty is presented, as the decision-maker knows the consequences of each action and how they are related to the other option.

## 2.2.4 Framing effect

One of the most important effects discovered during experimentation is the framing effect. This effect directly influences the reference point based on how the risky option is framed – negatively or positively. If the option is framed positively, it moves the reference point towards gain or towards loss, in the case of negative framing. In other words, specific framing of the available options has a psychological impact on the decision-maker's choices and can greatly influence if the person will be risk-averse or risk-prone. (Gvozdenović & Damnjanović, 2016) As the reference point represents the person's position before making a choice, framing the options in a positive or negative way moves the reference point into the gain or loss quadrants (Image 7), respectively. Because the decision-maker is not normatively rational, placing the reference point in the loss quadrant influences them to be more risk-prone, as the weight of loss becomes greater than the weight of gain. That is, people choose to "fix" their losses over gaining more than what they already have if placed in a frame of loss.

A simple example of the framing effect would be the following:

*Table 1 Framing effect question and answer example*

| Suppose you are given $500, but you have to pick between two options: | | |
|---|---|---|
| | Option A | Option B |
| **Positive framing** | 100% chance of **gaining** $100 | 50% chance of **gaining** $200 and 50% chance of **gaining** nothing |
| **Negative framing** | 100% chance of **losing** $100 | 50% chance of **losing** $200 and 50% chance of **losing** nothing |

## 2.2.5 Question design

Risk decision questions are consisted of two key components: surface structure and deep structure.

Deep structure implies the question itself which represents the situation in which the decision maker is placed, the certain option and the risky option (written in probabilistic terms), while surface structure represents anything which does not affect the meaning of the deep structure, such as the risk type, whether it is monetary, health based, etc. (Gvozdenović & Damnjanović, 2016)

The assumption then follows that the phrasing of the question, as well as the answers, will affect the decision-maker's *feeling* about the problem at hand and thus will influence their relationship with risk by placing them into the loss or gain quadrant. The expectation is that if the person is placed in a situation of loss, they will answer in a way which will move their reference point towards gain, while placing them in a situation of gain will influence them to pick the option which keeps the reference point in place or towards more gain.

Emotions play a significant role in decision-making and that is why the framing effect is so effective, as it directly affects choice based on what emotions are evoked within the person answering the questions.

## 3  Game design and implementation

The serious game is based in a 3D level, where the main character is played in first person. It is set in a fantasy medieval world and consists of a forest area and an enclosed town area. The development is divided into five major parts: questionnaire design, environment design, dialogue design, game logic and database implementation.

### 3.1  Questionnaire design

The game questionnaire[1] consists of seven questions (tasks) which are presented to the player by the non-playable characters upon interaction. All of the questions affect the player's character, a travelling adventurer, thus setting the player's reference point of gain or loss to be personal and, depending on the question, positive or negative. The questions are designed in both frames, only one of which appears in the game. Two versions of the game were built consisting of all of the seven tasks, so that both versions have positively and negatively framed questions shuffled and inversed from one another. Each question has two answers, whose outcome values are the same *objectively*, but are phrased in such a way so that there is always one certain and one risky option. The questionnaire is designed in a specific order; however, the gameplay does not insist on the player engaging with the tasks in the same way, the exception being that question 5 must be answered in order to unlock questions 6 and 7.

*Table 2 Framing label in both versions of the game of all of the questions*

| Version | Q1 | Q2 | Q3 | Q4 | Q5 | Q6 | Q7 |
|---|---|---|---|---|---|---|---|
| 1 | P | P | N | P | N | N | P |
| 2 | N | N | P | N | P | P | N |

#### Question 1

The player's character starts the game with 1 health point out of a maximum 250. As they approach the town, they meet a travelling salesman character who sells potions which would increase their health points. The player is presented with two options, one of which is represented in whole (certain) numbers and one in probabilistic terms (uncertain). Positive

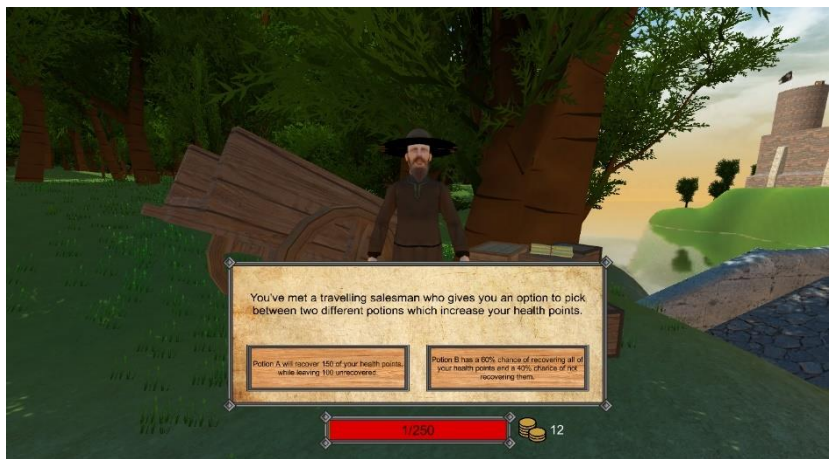

*Image 8 Game screenshot of question 1*

---
[1] The entire questionnaire is available in the Appendix of this thesis.

framing of this question indicates that the player's health points will be *recovered* or *unrecovered*, indicating a state of *healing* (gain) or *nothing* (neutral). Negative framing implies *damaged* (loss) and *undamaged* (neutral) health points.

### Question 2

The second question takes places at the gates of the city. The player is introduced to the current state of the town, which is closed off to visitors due to a bandit chase within the town walls. The player is presented with two options, one of which is paying a part of the entrance fee if they agree to help the bandit chase and the other attempting to convince the guard to pay less. Positive framing implies *receiving* gold coins back (gain) or *nothing* (neutral), while negative framing describes *paying* (loss) or *nothing* (neutral).

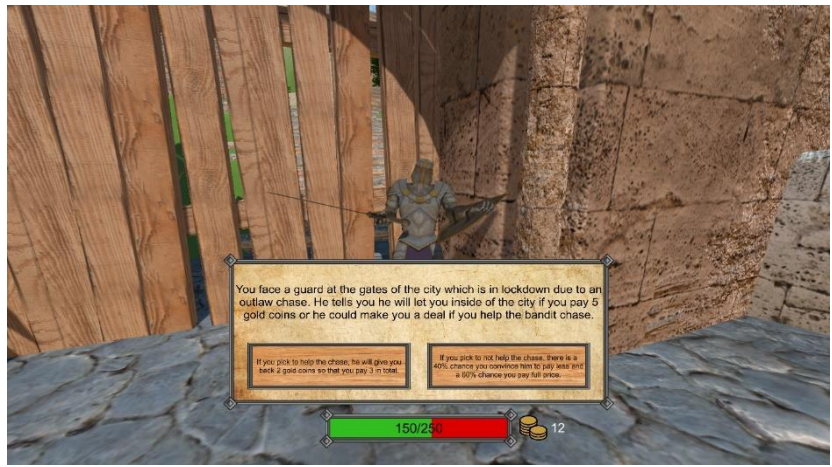
*Image 9 Game screenshot of question 2*

### Question 3

The third question is within the town walls. After entering the town, the player can roam within the walls and explore the area, while having the options of interacting with three different NPCs. This task is presented to the player by the town's butcher, to whom the player's character wants to sell their collectibles. Positive framing of the answer emphasizes how much money the player will make once they *sell* (gain) their valuables, while the negative focuses on how much they'll be *left* (loss) with if the butcher refuses to buy from them.

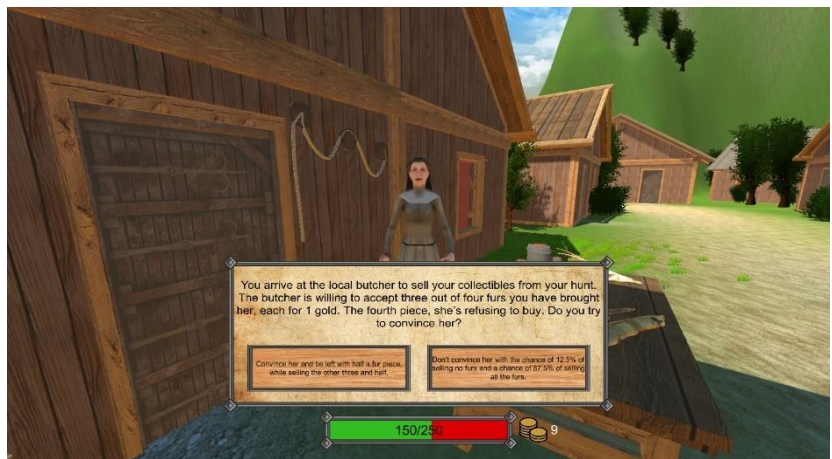
*Image 10 Game screenshot of question 3*

### Question 4

The fourth question takes place in the town's auction house and represents a gamble. The player is asked to pick between bidding a specific amount of gold or waiting to see if they win the auctioned item. Positive framing of the answers notes how much gold the player *remains* (gain) with, while the negative expresses *loss* (loss).

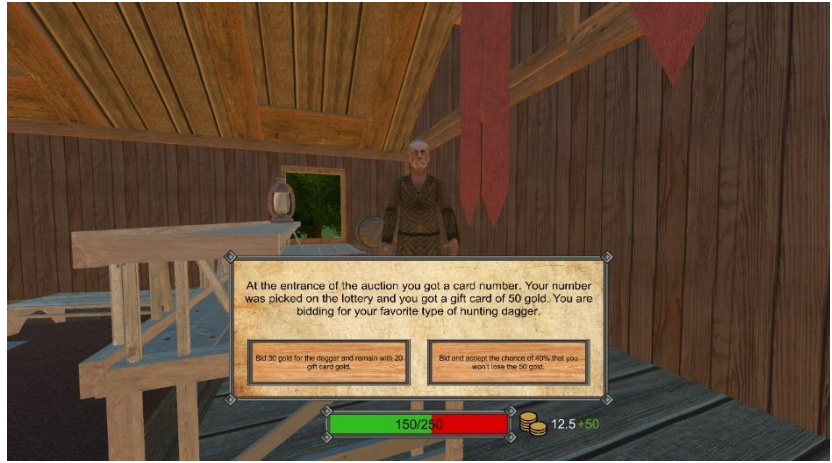

*Image 11 Game screenshot of question 4*

### Question 5

The fifth question puts the player in the position where they must pick how to attack the bandits. The answers directly affect the player's health points. The positive framing emphasizes the player's *defeat* (gain) of the bandits, while the negative explains the player's character getting *injured* (loss).

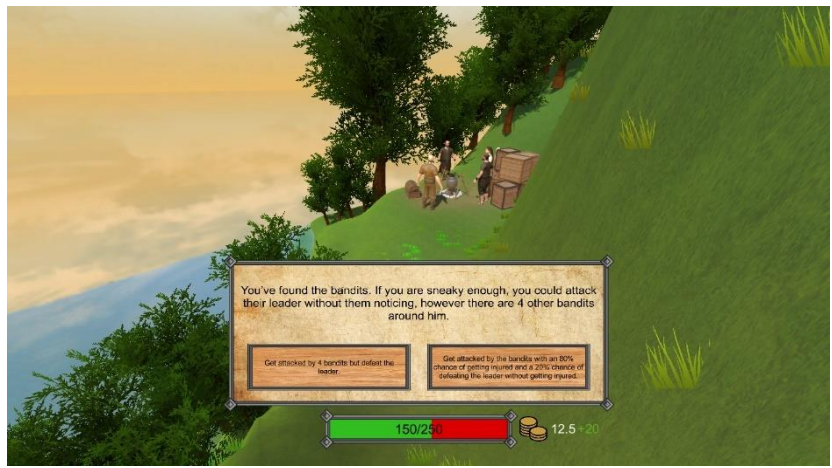

*Image 12 Game screenshot of question 5*

### Question 6

The sixth question, as well as the seventh, become available for answering once the player answers question five, since they are correlated thematically. The player arrives to the town's doctor after being injured by the bandits. The positive framing of the answers for this question is described by *pain relief* (gain), while the negative focuses on a huge operational risk with *even worse* (loss) consequences.

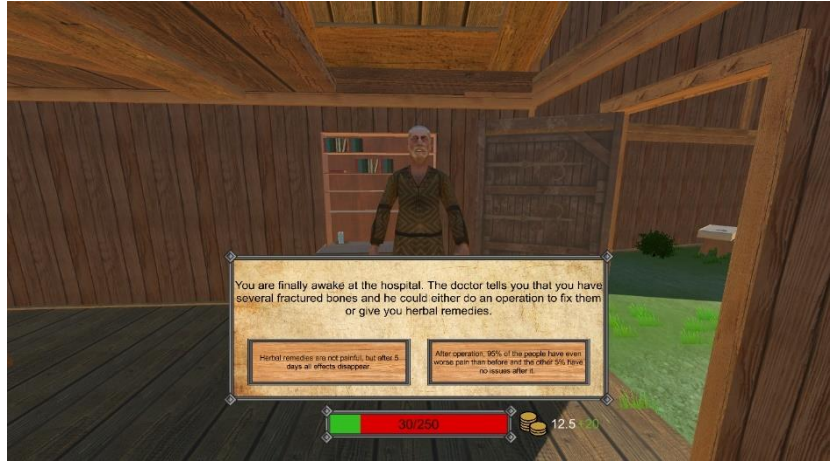

*Image 13 Game screenshot of question 6*

### Question 7

The seventh question is an adaptation of one of the most common prospect theory questions – the Asian disease problem. (Kahneman & Tversky, The Framing of Decisions and the Psychology of Choice, 1981) The task is to pick between two different plans of action, both of which have an effect on the town's population. The positive framing emphasizes *survival* (gain), while the negative focuses on *murder* (loss) of the people.

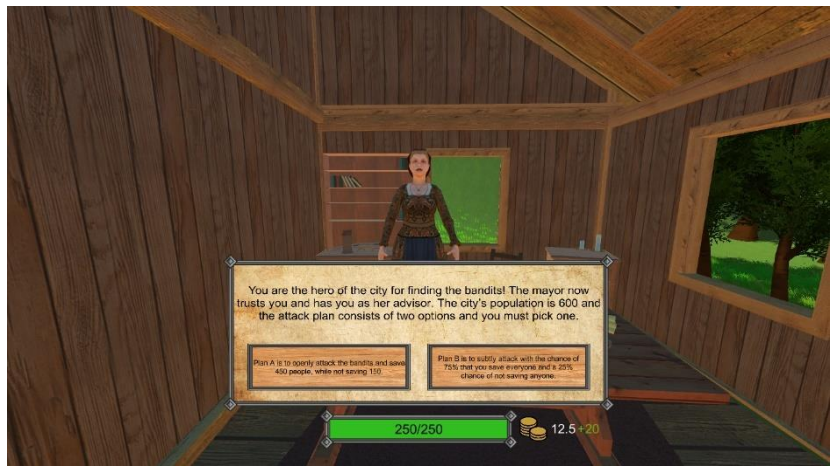

*Image 14 Game screenshot of question 7*

## 3.2 Environment design

### 3.2.1 Terrain design

Unity terrain is a plane game object which can be manipulated and modified using terrain tools. Terrain tools provide six different options for landscape creation and are available as a component as part of the terrain game object:

1. Create neighbouring terrain

This option creates interconnected plane game objects in a grid-like manner. The plane size is set to 1000x1000 [world units] by default, values which can be modified in the terrain settings options.

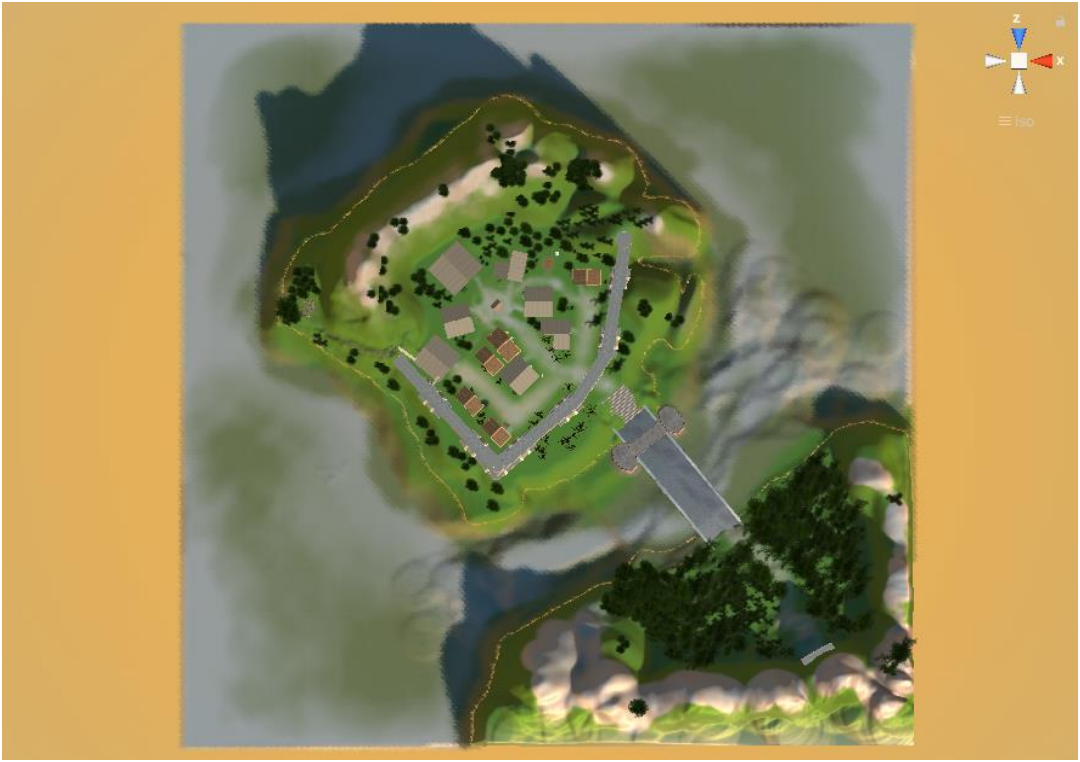

*Image 15 Topdown screenshot from the Unity editor of the game level terrain tile*

2. Paint terrain

Terrain painting consists of several terrain manipulation tools: raise/lower terrain (painting the terrain heightmap), paint holes (hide parts of the terrain), paint texture (apply textures), set height (set the heightmap to a specific value), smooth height (soften terrain features), stamp terrain (stamp a brush shape on top of the existing height map).

3. Paint trees

Trees are solid 3D game objects which are painted in a similar way to heightmaps. Before placing trees, a tree object must be assigned as a tree prefab[2] game object. Tree placement has several options. The most useful ones are brush size (dimensions of the placement range on the terrain) and tree density (average number of trees painted onto the brush size area). Unity uses optimization techniques such as billboarding in order to ensure good rendering performance.

---

[2] Nature Starter Kit – Asset store package which was used in this project for nature trees and grass details.

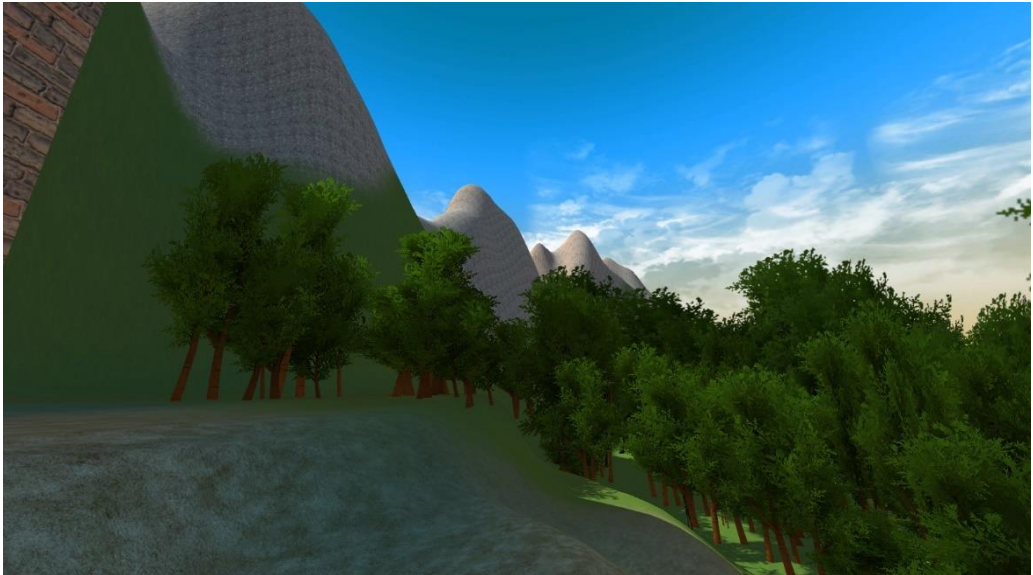
*Image 17 In game screenshot of the forest level area*

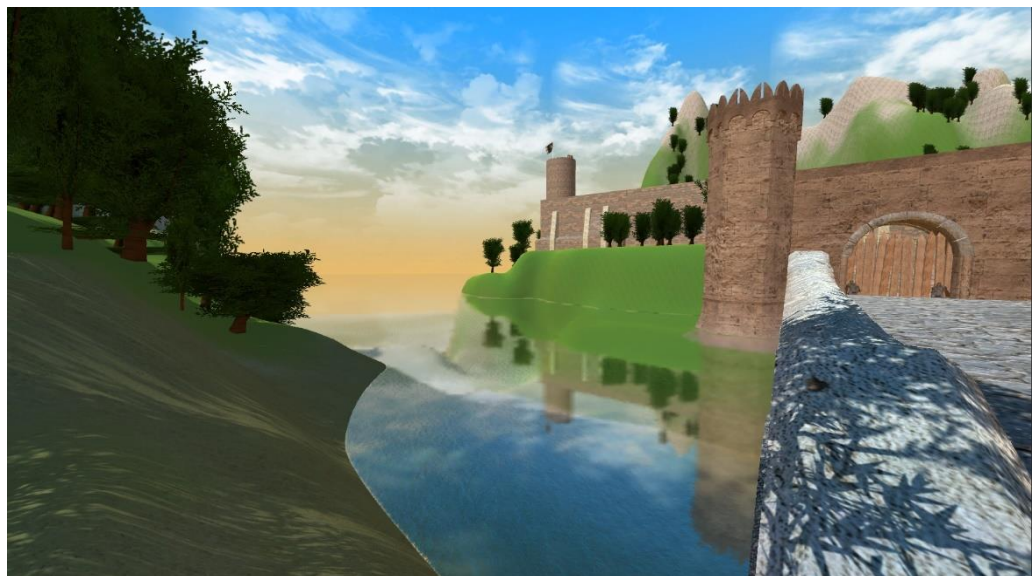
*Image 16 In game screenshot of the water from the bridge*

4. Paint details

Terrain details have similar options to trees, however they are not solid 3D objects, but textured primitive game objects which are 1 unit long or meshes, depending on the performance requirements. Enabling the billboard option for rendering makes it so that the textures automatically face the camera at all times, which makes rendering less costly and the gaming experience more lively.

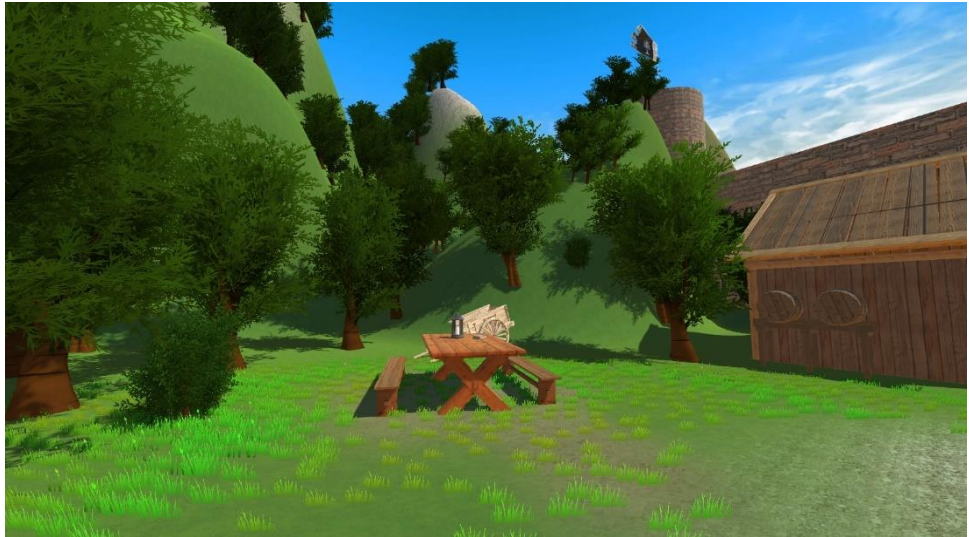
*Image 18 In game screenshot within the town walls*

5. Terrain settings

Terrain settings are primarily focused on rendering and shaping of the terrain planes and details, from basic terrain (plane management and restrictions), tree and details (size, density, rendering, etc.), texture resolutions, lighting and other settings.

### 3.2.2 Level and asset design

The level's props were gathered from different sources and used to create the game scene. This includes the standard assets, bridge, the bandit and villager characters, as well as the medieval props packs[3]. The world is constructed in such a way as to try to suspend disbelief but not make players take their environment too seriously. The overall design is stylized to fit the medieval theme, but it is not fantastical itself. Meaning, it does not include magical or fantasy elements, but rather sets the scene for the player visually to further grasp the player's attention while playing.

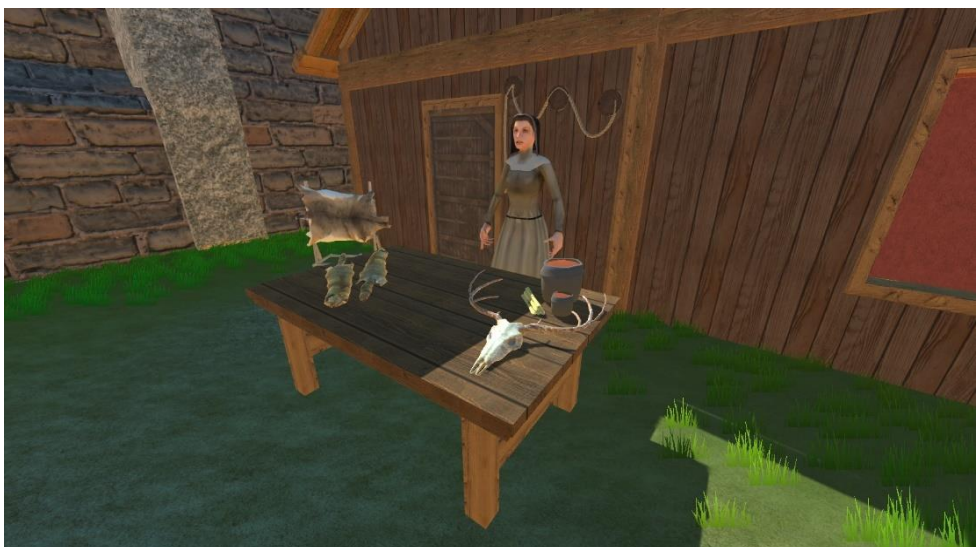
*Image 19 In game screenshot of the third task character, the butcher*

---

[3] Medieval props asset pack, Medieval props and Mega fantasy props pack

The main goals of the design were cohesion and feeling of clear purpose, so that the player is not distracted by the world, but rather uses it as a means to an end. The world has a secondary purpose to aid in setting the tone for the questions so that the hypothetical situations set in the defined problems become less hypothetical in their nature and more set on affecting the player's character themselves, thus removing the frame of the imaginary consequences of risky situations. There is a map with the overland section to guide the player's movement. The town itself contains a small selection of items and buildings and some NPCs, with whom the player interacts in order to answer the questions.

The player's starting position is designed so that it guides them to walk down a path towards the town gate and introduces the first question on the side of the road. Throughout the game, most of the questions can be answered in any order, giving the questionnaire an element of

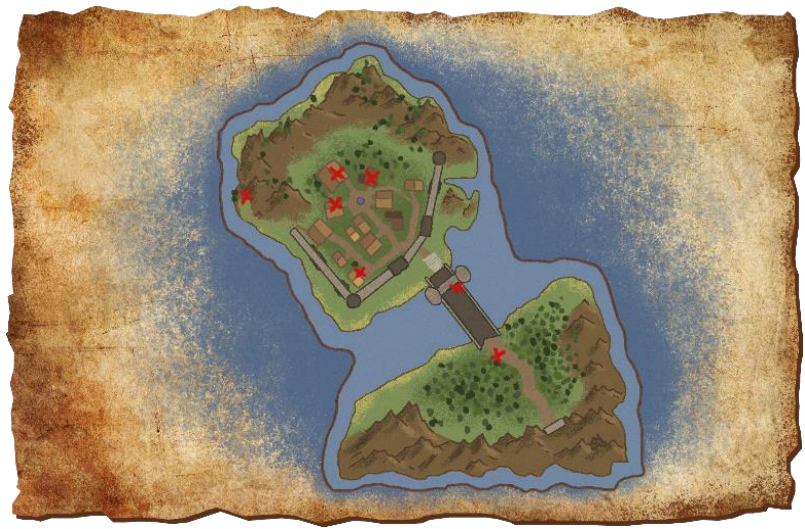

*Image 20 World map with marked question-giver positions*

randomness, with the exceptions being the following: in order to enter the town, the second question must be answered (the guard opening the gates) and before talking to the town doctor and mayor, the player must answer question five (finding the bandits). This greately affects the short story of the game, as the player can directly experience the consequences of their actions and decisions. This experience is also introduced by modifying the health points and gold coins amount of the player's character as a result of their decisions, for questions in the health and monetary domains, respectively. Due to the nature of prospect theory, the answers have the same numerical value and thus either of the two answers in each question arrives to the same outcome.

### 3.3 Dialogue design

#### 3.3.1 Dialogue script functions

Dialogue is the main part of this serious game, as it's the way players answer the questionnaire. This is completed through the dialogue class, dialogue manager and dialogue trigger scripts. A good dialogue sequence is needed to make the gameplay and story flow really well. It needs to keep the game fun, or at least interesting, by offering the player a seemless way of interacting with the NPCs. The dialogue is then played through in a user interface.

```csharp
public class Dialogue
{
    public string name;

    [TextArea(3, 10)]
    public string[] sentences;
}
```

The dialogue class object is instantiated in the dialogue trigger script and an interactable script, which are attached to each character object with whom the player interacts with. This makes it so that dialogue creation can easily be replicated on all of the characters, simply by changing the strings in the unity editor's components of the characters.

The interactable script is attached to the NPC as a component and it has two variables: a float radius and a boolean variable isInteractable. By default the boolean is set to true.

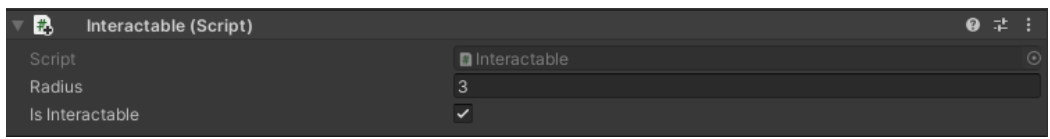
*Image 21 Unity editor interactable script component*

The radius variable is used for debugging, as it displays a wired sphere around the object to display its interactability, while the boolean is used to mark if the object can be interacted with or not and its value is changed when the player answers the specific character's question and solves the task. This makes it so that the player can answer the question only once, as well as counting the number of questions which have been answered already.

The sentences are stored in an array of strings and the strings denote, in order: the question, first answer, second answer and continuation of dialogue text after the question has been answered. The name variable is used to denote the NPCs name or role.

Dialogue trigger script has two additional functions, which are used to initiate the dialogue, and continue it upon answering. These void functions are used as triggers on the answer and continue buttons on click and they call upon the dialogue manager and two of its functions, passing in the existing dialogue object as its parameter.

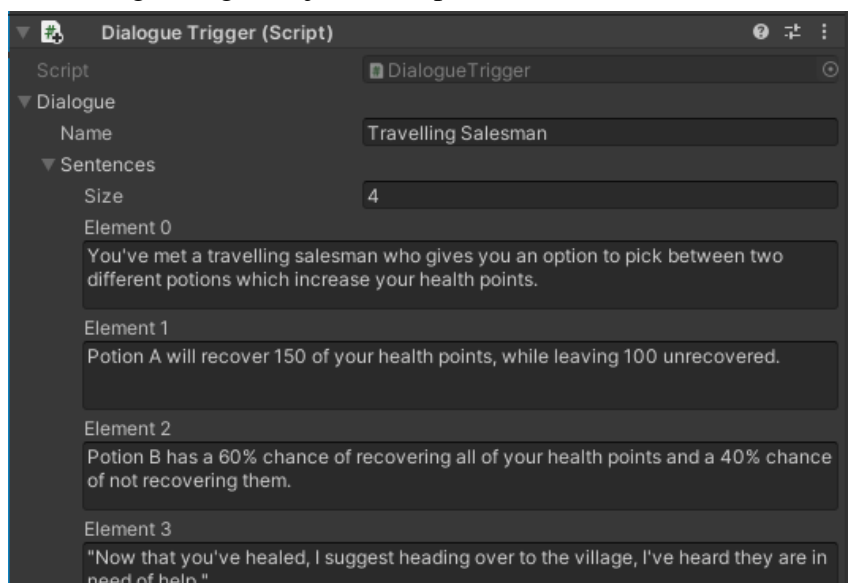
*Image 22 Unity editor dialogue trigger script component*

```
public void TriggerDialogue()
{
    FindObjectOfType<DialogueManager>().StartDialogue(dialogue);
}

public void ContinueDialogue()
{
    FindObjectOfType<DialogueManager>().Continue(dialogue);
}
```

Dialogue manager script takes care of the majority of the dialogue actions.

```
public void StartDialogue(Dialogue dialogue)
{
    Debug.Log("Starting conversation with " + dialogue.name);
    dialoguePanel.SetActive(true);

    continueButton.SetActive(false);
    buttonOne.SetActive(true);
    buttonTwo.SetActive(true);

    sentences.Clear();

    dialogueText.text = dialogue.sentences[0];

    buttonOne.transform.GetChild(0).gameObject.GetComponent<Text>().text = dialogue.sentences[1];

    buttonTwo.transform.GetChild(0).gameObject.GetComponent<Text>().text = dialogue.sentences[2];
}
```

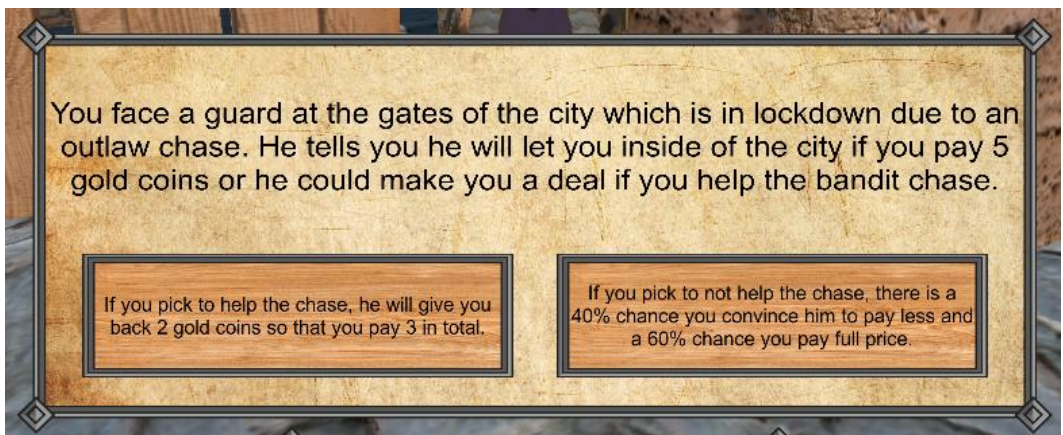

*Image 23 In game screenshot of the second task dialogue panel*

The start dialogue function displays the question panel which contains the text boxes for the problem text and the answer buttons, as well as clearing the sentences queue and placing the texts from the dialogue object's sentences array into their corresponding fields.
While the dialogue panel is active, the first person controller's camera movement is disabled and the camera position is rotated towards the non-playable character's center in order face the player towards them, as well as rotating the character towards the camera, i.e. the player character. Since the mouse arrow is disabled by default while the character is moving through the environment, starting the dialogue displays the arrow so that the user can interact with the panel.

```csharp
public void Continue(Dialogue dialogue)
{
    continueButton.SetActive(true);
    buttonOne.SetActive(false);
    buttonTwo.SetActive(false);

    dialogueText.text = dialogue.sentences[3];
}
```

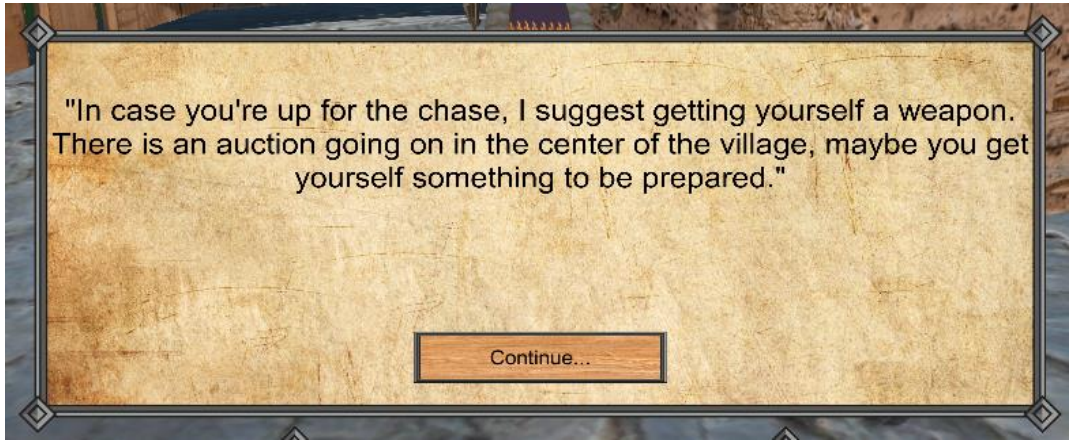

*Image 24 In game screenshot of the task dialogue panel after answering the question*

The continue function deactivates the answer buttons and displays the continue button, which displays the continuation text taken from the dialogue object's string array. The continue button closes the panel, unlocks the camera movement, hides the mouse arrow and marks the isInteractable boolean from the interactable script as false, thus closing the entire interaction circle of the selected character.

### 3.3.2 First person controller script

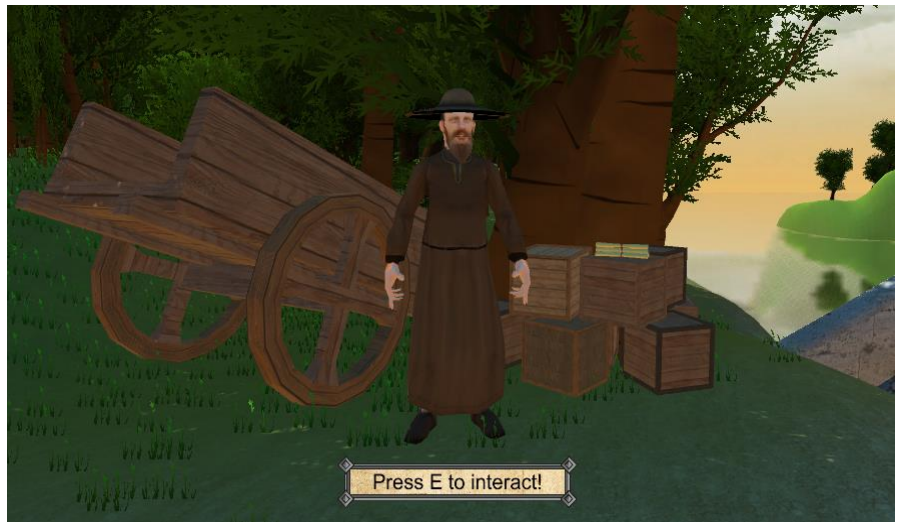

*Image 25 In game screenshot of an interactable character with the interaction notification displayed*

In order for the player to interact with the characters, ray casting is used. Physics ray casting casts a ray towards a specified direction and returns true if the ray collides with an object up to a specified distance, else it returns false if there is no collision. Each interactable character has a capsule collider around them with which the rays collide when the character is close and facing the characters. The first person controller script contains a variable focus, which is an interactable class object and is set to null if no rays are casted towards interactable game objects which have the interactable script component attached to them.

```csharp
if (focus == null)
{
    Ray ray = m_Camera.ScreenPointToRay(Input.mousePosition);
    RaycastHit hit;

    if (Physics.Raycast(ray, out hit, 20))
    {
        target = hit.collider.gameObject;
        NPCrotationDefault = target.transform.rotation;

        Interactable interactable = hit.collider.gameObject.GetComponent<Interactable>();

        if (interactable != null && interactable.isInteractable)
        {
            interactionNotice.SetActive(true);

            if (CrossPlatformInputManager.GetButtonDown("Interact"))
            {
                SetFocus(interactable);

                interactionNotice.SetActive(false);

                Debug.Log("Focused on target!");

                Vector3 npcLocation = target.GetComponent<CapsuleCollider>().transform.position;

                npcLocation = new Vector3(npcLocation.x, target.GetComponent<CapsuleCollider>().bounds.center.y, npcLocation.z);

                transform.GetChild(0).gameObject.transform.LookAt(npcLocation, Vector3.up);

                target.transform.LookAt(transform);

                if (target.GetComponent<Animator>())
                {
                    target.GetComponent<Animator>().SetTrigger("IsInteracting");
                }
            }
        } else
        {
            interactionNotice.SetActive(false);
        }
    } else
    {
        interactionNotice.SetActive(false);
    }

    if (CrossPlatformInputManager.GetButtonDown("Cancel") && focus != null)
    {
        Debug.Log("Removed focus");

        RemoveFocus();
    }
}
```

Once the character ray casts an interactable object, whose isInteractable boolean variable is set to true, the player is shown a panel to indicate the availability of interaction with that character.

If the first person controller's focus variable is set to null, ray casting is performed to check for interactable characters. Interaction becomes available if the ray is cast on an object with a capsule collider and an interactable script component which has the isInteractable variable set to true. Once the player presses the interact button (E key), the SetFocus function is called, passing the ray casted's interactable component as a parameter.

```
void SetFocus(Interactable newFocus)
{
    focus = newFocus;

    mouseLookEnabled = false;

    MouseLook.m_cursorIsLocked = false;

    focus.gameObject.GetComponent<DialogueTrigger>().TriggerDialogue();
}
```

This function assigns the interactable object to the focus variable, disables camera movement, unlocks the mouse arrow and calls upon the trigger dialogue function from the dialogue triggers script. Alternatively, the RemoveFocus function does the opposite. Once the continue button is pressed in the dialogue panel or the player presses the cancel button (Escape key), this function is called to close the dialogue panel and allow camera movement, as well as hiding the mouse arrow once again so that the character can freely move around again.

```
public void RemoveFocus()
{
    mouseLookEnabled = true;

    MouseLook.m_cursorIsLocked = true;

    focus.gameObject.transform.rotation = NPCrotationDefault;

    FindObjectOfType<DialogueManager>().dialoguePanel.SetActive(false);

    if (target.GetComponent<Animator>())
    {
        target.GetComponent<Animator>().SetTrigger("IsInteracting");
    }

    focus = null;
}
```

The continue dialogue function is attached to the continue button and it activates the dialogue trigger's corresponding function.

```
public void ContinueDialogue()
{
    focus.gameObject.GetComponent<DialogueTrigger>().ContinueDialogue();
}
```

## 3.4 Game logic

The player movement is handled by the [standard assets package](#)'s first person controller imported in Unity, a script to which code was additionally added in the previous section for character interaction and dialogue management.

The game manager script is the core of this video game. It handles database connection, user personal information and answers, task logic, as well as variable and environment alteration as a consequence of the player's choices in the game world. Most of the crucial game objects are added to the game manager script within the Unity editor.

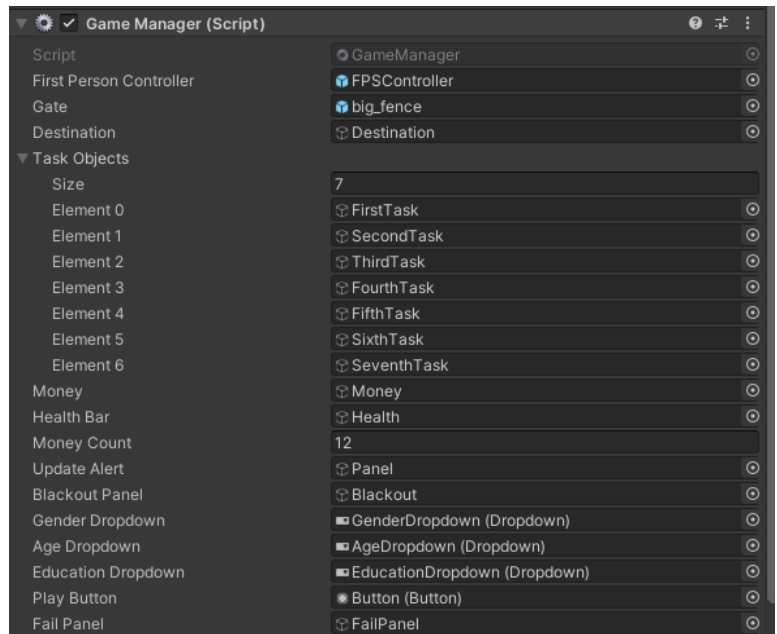

*Image 26 Game manager script component*

### 3.4.1 Task logic

All of the seven tasks within the game have consequences on the player and the world around them as they move through the world an engage with the environment by answering the questions given by the NPCs.

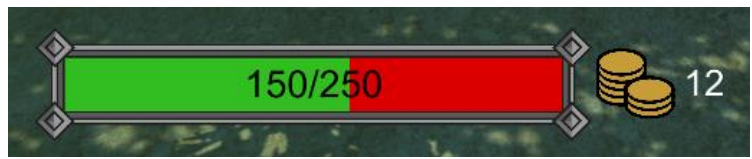

*Image 27 In game screenshot of the main character's health points and gold coins*

The player's initial state is defined by 1 health point and 12 gold coins. In order to track and mark the tasks as solved, the game manager keeps an array called TaskObjects of game objects which are added within the Unity editor and it contains the game objects which have the Task class attached to them. The task class is a simple component of each object which has a boolean variable with the information if the task is solved or not.

```csharp
public class Task : MonoBehaviour
{
    public bool isSolved = false;
}
```

Additionally, the game manager keeps count of how many tasks have been solved. By pressing either of the two answer buttons, the user's answer is saved as an integer 1 or 2, first or second answer respectively, the task is marked as solved which disables the intractability of the character and increases the solved tasks count. The markSolved function also implements the task consequences. The following code snippets are part of that function and each task is explained separately, but before the tasks are implemented, the script increases the solvedCount integer variable, extracts the focused target's name, marks the task as solved and sets the intractability to false.

After each question is answered, the player is shown a text alert indicating the consequence of their choice for 3 seconds using coroutines. Coroutines are functions which have the ability to pause execution, return the control to Unity, but continue where they had left off in the previous frame. This makes it so that coroutine functions (or IEnumerators) can be executed in parallel with Unity's scripts without visible interruptions, but more importantly make timers available. The yield return function is where the IEnumerator pauses and continues each frame.

```csharp
IEnumerator ShowAlert(string textToDisplay)
{
    updateAlert.transform.GetChild(0).gameObject.GetComponent<Text>().text = textToDisplay;
    updateAlert.SetActive(true);
    yield return new WaitForSeconds(3);
    updateAlert.SetActive(false);
}
```

The ShowAlert coroutine is used after each task and it displays the text in the alert panel, a string which is passed to the function as a parameter.

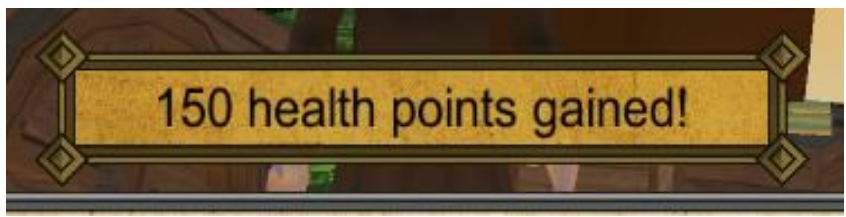

*Image 28 An example of the alert notification displaying an increase in health points as a consequence of the first task*

```csharp
solvedCount++;

Interactable focus = firstPersonController.GetComponent<FirstPersonController>().focus;

focus.transform.parent.gameObject.GetComponent<Task>().isSolved = true;
focus.GetComponent<Interactable>().isInteractable = false;

string taskNumber = focus.transform.parent.name;
```

The task number variable is used to check which task is currently in focus and in respect to that code is executed.

```
if (System.String.Equals(taskNumber, "FirstTask"))
{
    healthBar.GetComponent<RawImage>().texture = Resources.Load<Texture2D>("healthBarFilled");
    healthBar.transform.GetChild(0).gameObject.GetComponent<Text>().text = "150/250";

    StartCoroutine(ShowAlert("150 health points gained!"));
}
```

After completing the first task, the health bar texture and string are changed to display 150 health points visually and numerically.

```
else if (System.String.Equals(taskNumber, "SecondTask"))
{
    moneyCount -= 3f;
    money.transform.GetChild(0).gameObject.GetComponent<Text>().text = (moneyCount).ToString();
    moveGate();

    StartCoroutine(ShowAlert("3 gold coins lost!"));
}
```

```
public void moveGate()
{
    float speed = 5000;
    Transform pivotPoint = gate.transform.GetChild(0).gameObject.transform;

    gate.transform.RotateAround(pivotPoint.position, Vector3.up, -speed * Time.deltaTime);
    gate.GetComponent<MeshCollider>().enabled = false;
}
```

Completion of the second tasks decreases the amount of gold the player has and opens the gates to the city by pivoting the gate game object around an empty game object placed on the gate's edge and disabling the mesh collider component of the object.

```
else if (System.String.Equals(taskNumber, "ThirdTask"))
{
    moneyCount += 3.5f;
    Debug.Log(moneyCount);
    money.transform.GetChild(0).gameObject.GetComponent<Text>().text = (moneyCount).ToString();

    StartCoroutine(ShowAlert("3.5 gold coins gained!"));
}
```

The third task does a very similar code sequence to the second, only it adds gold instead of subtracting it.

```csharp
else if (System.String.Equals(taskNumber, "FourthTask"))
{
    money.transform.GetChild(1).gameObject.GetComponent<Text>().text = "+20";

    StartCoroutine(ShowAlert("You've won the dagger!"));
}
```

The fourth task is monetary, much like the second and third, the only difference being it displays bonus auction gold coins which are displayed next to the existing coins the player has.

```csharp
else if (System.String.Equals(taskNumber, "FifthTask"))
{
        taskObjects[5].transform.GetChild(0).gameObject.GetComponent<Interactable>(
        ).isInteractable = true;

        taskObjects[6].transform.GetChild(0).gameObject.GetComponent<Interactable>(
        ).isInteractable = true;

        StartCoroutine(FadeToBlack());

        StartCoroutine(ShowAlert("You've been injured!"));

        healthBar.GetComponent<RawImage>().texture =
        Resources.Load<Texture2D>("healthBarKindaFilled");
        healthBar.transform.GetChild(0).gameObject.GetComponent<Text>().text =
        "30/250";
}
```

The completion of the fifth task has an important segment, which is setting the intractability of the sixth and seventh tasks to true, as the completion of this task is required in order for the last two tasks to become available. The sixth and seventh task objects are accessed from the taskObjects array which calls upon the Interactable script component of the objects and sets the isInteractable boolean to true. Once the fifth task is completed, another coroutine is called which fades the screen to black and back to create the effect of being injured by the bandits and moving the player's location to the sixth task (the town doctor), as well as decreasing the player's health points.

```csharp
IEnumerator FadeToBlack()
{
    while (blackoutColor.a < 1)
    {
        fadeAmount = blackoutColor.a + (1 * Time.deltaTime);

        blackoutColor = new Color(blackoutColor.r, blackoutColor.g,
blackoutColor.b, fadeAmount);
        blackoutPanel.GetComponent<Image>().color = blackoutColor;

        yield return null;
    }

    firstPersonController.GetComponent<FirstPersonController>().enabled = false;
    firstPersonController.transform.position = new Vector3(455, 212, 668);
    firstPersonController.GetComponent<FirstPersonController>().enabled = true;
```

```
    while (blackoutColor.a > 0)
    {
    fadeAmount = blackoutColor.a - (1 * Time.deltaTime);

    blackoutColor = new Color(blackoutColor.r, blackoutColor.g, blackoutColor.b,
fadeAmount);
    blackoutPanel.GetComponent<Image>().color = blackoutColor;

    yield return null;
    }
}
```

Once the character interacts with the doctor and completes the sixth task, the health bar is updated in the same way as in the first task, making the player's health points completely full.

```
else if (System.String.Equals(taskNumber, "SixthTask"))
{
    healthBar.GetComponent<RawImage>().texture =
Resources.Load<Texture2D>("healthBarFull");
    healthBar.transform.GetChild(0).gameObject.GetComponent<Text>().text =
"250/250";

    StartCoroutine(ShowAlert("Healing steadily!"));
}
```

### 3.4.2 User logic

User or player information is stored in a UserAnswer class object and it includes information on the player's gender, age, education background and the answers they picked.

```
public class UserAnswer
{
    public string gender;
    public int age;
    public string education;
    public List<int> answers;
}
```

Gender, age and education variables are set through dropdowns when the game starts and the player is given the option of leaving the fields empty or selecting from the available options. By default, the game manager fills the answers list with seven zeroes, indicating that no tasks have been solved and making the answers list of length 7, which is later used for easier indexing of the solved tasks. This is done as an alternative to using a dictionary, as it makes it so that the order of answering doesn't matter as the list answers are sorted like written in the questionnaire. Both of the answer buttons have their corresponding answer storage functions where they place the answers 1 or 2 in the user answers list at the specific index *task number - 1*. The following snipped is the implemented function for the first answer button and the second answer button only differentiates from this at the last line, where the inserted answer is 2, instead of 1.

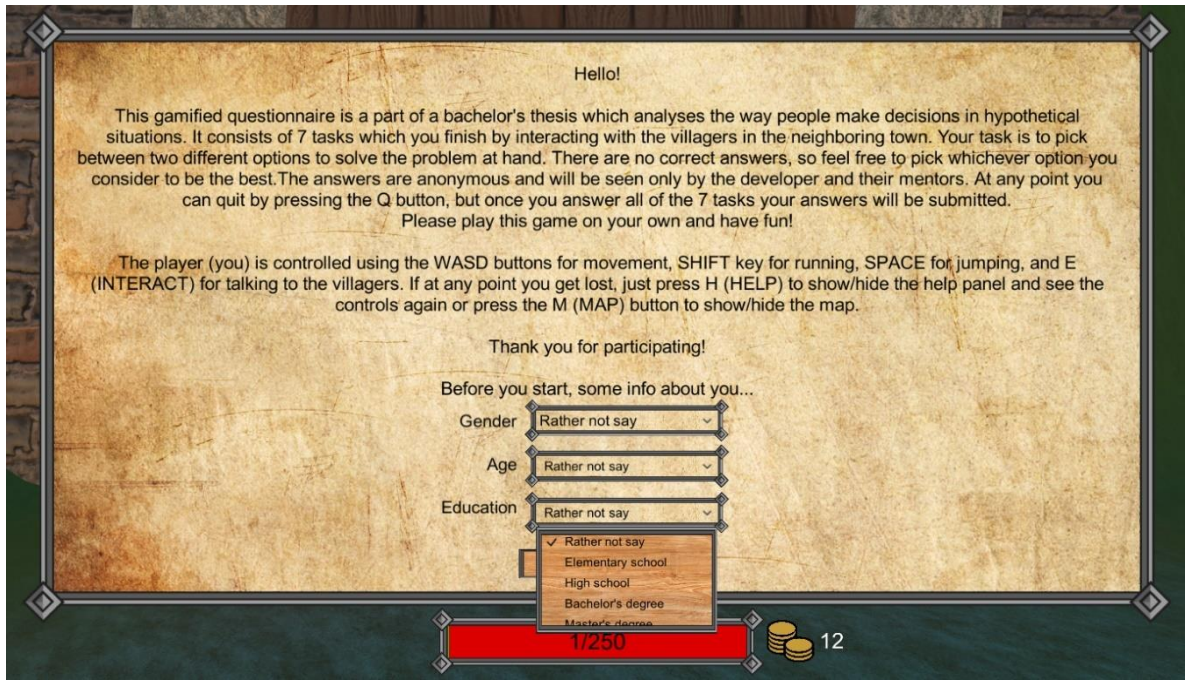

*Image 29 In game screenshot of the welcome panel where the user inputs their information*

```
public void insertAnswerOne()
{
    Interactable focus = 
firstPersonController.GetComponent<FirstPersonController>().focus;
    string taskNumber = focus.transform.parent.name;

    switch (taskNumber)
    {
        case "FirstTask":
            taskNum = 0;
            break;
        case "SecondTask":
            taskNum = 1;
            break;
        case "ThirdTask":
            taskNum = 2;
            break;
        case "FourthTask":
            taskNum = 3;
            break;
        case "FifthTask":
            taskNum = 4;
            break;
        case "SixthTask":
            taskNum = 5;
            break;
        case "SeventhTask":
            taskNum = 6;
            break;
    }

    activeUser.answers[taskNum] = 1;
    checkIfAllSolved();
}
```

The checkIfAllSolved function checks if all seven tasks have been completed from the game manager's solvedCount variable, as well as checking if the answers have already been saved into the database, in order to avoid duplicate entries. If the answers have not been saved and all tasks are completed, the user object is sent to the database.

```
public void checkIfAllSolved()
{
    if (solvedCount == 7 && !savedAnswers)
    {
        SaveAnswersToDatabase(activeUser);
    }
}
```

## 3.5   Database implementation

### 3.5.1   MongoDB

MongoDB is a NoSQL document-oriented database program which stores data in JSON-like document formats with optional schemas. The document fields can vary in each document and the structure can be changed over time, making the data management and modification more flexible and convenient. Due to the nature of this game's data which is stored, a NoSQL database was selected as it is a fast and convenient way to store data which would have otherwise been stored in a single relational database table.

### 3.5.2   Unity and MongoDB

Offline video games often store their data locally on the player's computer. However, for the purpose of this thesis, the user data of all players is required to be stored in a single place in order to statistically analyze the questionnaire answers. The data is stored on the MongoDB Cloud, specifically MongoDB Atlas, which is a fully managed cloud database developed for modern applications. (MongoDB Cloud, 2021)
Atlas offers a free database cluster with 512MB of storage, shared RAM, end-to-end encryption, a maximum of 100 connections, 100 databases and 500 collections. These specifications are more than enough for the purposes of this project, as each document within a collection is no more than 150B. The database *ProspectTheory* has two collections, *AnswersV1* and *AnswersV2*, for the first and second version of the game (differentiating in framing types of each question), respectively.

The document schema of each user is the following:

```
{
    "_id" : "$oid",
    "gender" : "$string",
    "age" : "$numberInt",
    "education" : "$string",
    "answers":[
        "$numberInt",
        "$numberInt",
        "$numberInt",
        "$numberInt",
        "$numberInt",
        "$numberInt",
        "$numberInt"
    ]
}
```

In order to connect to the database, Unity requires MongoDB C# drivers[4] placed in the project's Plugins folder. On the Atlas side, a database user is created through the web interface with the admin role and the user is assigned a connection string. This string is used within the code to connect to the database and perform CRUD operations on the database collections.

```
try {
    var mongoUrl = new MongoUrl(connectionString);
    client = new MongoClient(mongoUrl);
    database = client.GetDatabase("ProspectTheory");
    collection = database.GetCollection<BsonDocument>("AnswersV2");

    client.ListDatabaseNames();
} catch (Exception e) {
    playButton.interactable = false;
    failPanel.SetActive(true);
}
```

The database connection code is enclosed within a try-catch block in order to deny the player gameplay if the game fails to connect. This is done so that the player cannot play the game without database access, as the data they entered will not be saved. The code example above is connecting to the atlas ProspectTheory database, collection AnswersV2. Game version 1 is different from this snipped only on the line of the collection name, as it is set to AnswersV1.

The collection type is a BsonDocument (Binary JSON), which is a binary-encoded serialization of JSON-like documents and an object model used when the data is free form with different types and structures of the information within the data object, such as the user object in this project, which contains integers, strings and lists.

---

[4] MongoDB C# drivers are available on the following github repository.

- 36 -

```csharp
public async void SaveAnswersToDatabase(UserAnswer user)
{
    BsonArray answersArray = new BsonArray();

    foreach (var a in user.answers)
    {
        answersArray.Add(a);
    }

    var document = new BsonDocument
    {
        { "gender", user.gender },
        { "age", user.age },
        { "education", user.education },
        { "answers", answersArray }
    };

    await collection.InsertOneAsync(document);

    savedAnswers = true;
}
```

Async functions in C# allow code to be paused and resumed, much like previously mentioned coroutines. The difference is that these types of methods return data. This function takes a user object as a parameter and converts it to a BsonDocument in order to insert it into the database. Asynchronization is done in order to prevent interruption of the code execution while playing the game and ensuring that the document is inserted into the database correctly. The first three fields are inserted into the BsonDocument format directly, while the answers list is converted into a BsonArray before insertion.

Once the answers are saved, the game manager's savedAnswers boolean is set to true and the game is finished.

# 4 Conclusion

The development of this serious game has led many new similar ideas to arise, both through testing and feedback from psychologists whose primary field is decision-making processes. Both versions of the game combined was finished by 25 people in total from various backgrounds and most have noted that playing the game and being engaged with the virtual world has led them to setting aside the fact that they are completing a psychological questionnaire. This is worth noting because it is one of the main hypotheses behind the development. Creating a virtual world with which the players can directly interact with and face tangible consequences of their decisions brings the hypothetical situations of the prospect theory questions closer to the respondent. If the players can interact with and manipulate the virtual world, then it is much easier for them to empathize with it and their character, and thus, the assumption is that the answers represent a more realistic image of the player's decision making.

This hypothesis proposes a further research question which is to compare the answers of people who have answered the same questionnaire on paper and through playing the game. The comparison would show if the simulation of a prospect theory questionnaire has an effect on decision-making.

Modifications of the answers themselves could also offer a different insight into the decision-making of the players, by showing parts of the answers and leaving up to the player to conclude the (un)certainty of their choices or by shuffling the answers on the buttons within the game, i.e. not showing the certain option always in the place of the first answer. An example of the partial answer is the following:

Potion A will recover 150 of your health points, while leaving 100 unrecovered.
*will become*
Potion A will recover 150 of your health points.
*or alternatively*
Potion A will leave 100 of your health points unrecovered.

Additionally, a timer can be included within the game in order to measure the amount of time the user takes in order to answer each question.

Development of serious games for psychological questionnaires is not common practice, however it shows great potential as it engages the respondents on a deeper level with the task at hand, only it requires using a computer as well as some previous knowledge of interacting with video game worlds. Potential workarounds for these two problems would be development of serious games for mobile platforms and including detailed but short tutorials or maximizing intuitivity and simplification of the human-computer interaction within the game.

# 5 References

1. Allais, M. (1953). Le Comportement de l'Homme Rationnel devant le Risque: Critique des Postulats et Axiomes de l'Ecole Americaine. *Econometrica*, 503-546.
2. Cummings, J. J., & Bailenson, J. N. (2016). How Immersive Is Enough? A Meta-Analysis of the Effect of Immersive Technology on User Presence. *Media Psychology, 19*, 272-309.
3. Damnjanović, K. M. (2014). *Cognitive factors of the framing effect in decision-making tasks.* Belgrade: University of Belgrade.
4. Deterding, S., Dixon, D., Khaled, R., & Nacke, L. (2011). From Game Design Elements to Gamefulness: Defining "Gamification". *Proceedings of the 15th International Academic MindTrek Conference: Envisioning Future Media Environments, MindTrek*, 9-15.
5. Elgar, E. (1998). Expected Utility Theory. In E. Elgar, *The Handbook of Economic Methodology* (pp. 171-178). Cheltenham: Edward Elgar Pub.
6. Gvozdenović, V., & Damnjanović, K. (2016). Influence of the probability level on the framing effect in reference point of loss. *Primenjena psihologija*.
7. Haas, J. (2014). *A History of the Unity Game Engine.* Worcester polytechnic institute.
8. Kahneman, D. (2011). *Thinking, fast and slow.* Farrar, Straus and Giroux.
9. Kahneman, D., & Tversky, A. (1979). Prospect Theory: An analysis of Decision under Risk. *Econometrica*, 263-291.
10. Kahneman, D., & Tversky, A. (1981). The Framing of Decisions and the Psychology of Choice. *Science*, 453-458.
11. McLeod, S. A. (2018). *Questionnaire: definition, examples, design and types.* Retrieved from Simply Psychology: https://www.simplypsychology.org/questionnaires.html
12. Megha, P., Nachammai, L., & Senthil Ganesan, T. M. (2018). 3D Game Development using Unity Game Engine. *International Journal of Scientific & Engineering Research*.
13. Messaoudi, F., Simon, G., & Ksentini, A. (2015). Dissecting Games Engines: the Case of Unity3D. *ACM/IEEE Netgames : The 14th International Workshop on Network and Systems Support for Games.* Zagreb.
14. *MongoDB Cloud*. (2021). Retrieved from mongoDB: https://www.mongodb.com/cloud
15. Wagner, B. (2021, 8 23). *A tour of the C# language*. Retrieved from Microsoft: https://docs.microsoft.com/en-us/dotnet/csharp/tour-of-csharp/

# 6 Appendix

A video demonstration of the gameplay is available on [this link](this link).
The script source code is available on [github](github).

## 6.1 Questionnaire

**1. You meet a travelling salesman who gives you an option to pick between two different potions which increase your health points (max 250).**

### POSITIVE FRAMING

1. Potion A will recover 150 of your health points, while leaving 100 unrecovered.
2. Potion B has a 60% chance of recovering all of your health points and a 40% chance of not recovering them.

### NEGATIVE FRAMING

1. Potion A will leave you with 100 damaged health points and 150 undamaged health points.
2. Potion B has a 40% chance of damaging your health points and a 60% chance of not damaging them.

**2. You face a guard at the gates of the city which is in lockdown due to an outlaw chase. He tells you he will let you inside of the city if you pay 5 gold coins or he could make you a deal if you help the bandit chase.**

### POSITIVE FRAMING

1. If you pick to help the chase, he will give you back 2 gold coins so that you pay 3 in total.
2. If you pick to not help the chase, there is a 40% chance you convince him to pay less and a 60% chance you pay full price.

### NEGATIVE FRAMING

1. If you pick to help the chase, you pay 5 gold and he gives you back 2.
2. If you pick to not help the chase, there is a 60% chance that you won't pay full price and a 40% chance you will pay 5 gold.

**3. You arrive at the local butcher to sell your collectibles from your hunt. The butcher is willing to accept three out of four furs you have brought her, each for 1 gold. The fourth piece, she's refusing to buy. Do you try to convince her?**

    **POSITIVE FRAMING**

1. Convince her and get 3.5 gold in total for the furs, the last one being half the price.
2. Don't convince her with the chance of 87.5% of selling all of the furs and a 12.5% chance of selling no furs.

    **NEGATIVE FRAMING**

1. Convince her and be left with half a fur piece, while selling the other three and half.
2. Don't convince her with the chance of 12.5% of selling no furs and a chance of 87.5% of selling all the furs.

**4. At the entrance of the auction you got a card number. Your number was picked on the lottery and you got a gift card of 50 gold. You are bidding for your favorite type of hunting dagger.**

    **POSITIVE FRAMING**

1. Bid 30 gold for the dagger and remain with 20 gift card gold.
2. Bid and accept the chance of 40% that you won't lose the 50 gold.

    **NEGATIVE FRAMING**

1. Bid for the dagger and lose 30 gift card gold, while not losing 20.
2. Bid and accept the chance of 60% that you lose all of your gold, with the chance of 40% of not losing any.

**5. You've found the bandits. If you are sneaky enough, you could attack their leader without them noticing, however there are 4 other bandits around him.**

    **POSITIVE FRAMING**

1. Attack and defeat the leader, while alerting the other 4.
2. Attack with the chance of 20% that you defeat all of the bandits and an 80% chance of not defeating any of them.

    **NEGATIVE FRAMING**

1. Get attacked by 4 bandits but defeat the leader.
2. Get attacked by the bandits with an 80% chance of getting injured and a 20% chance of defeating the leader without getting injured.

**6. You are finally awake at the hospital. The doctor tells you that you have several fractured bones and he could either do an operation to fix them or give you herbal remedies.**

    **POSITIVE FRAMING**

1. Herbal remedies will relieve the pain for 5 days, after which your limbs will become painful again.
2. Take the operation which is very risky and 5% of the people who undertake it have no issues after it, while 95% of the people have even worse pain than before.

    **NEGATIVE FRAMING**

1. Herbal remedies are not painful, but after 5 days all effects disappear.
2. After operation, 95% of the people have even worse pain than before and the other 5% have no issues after it.

**7. You are the hero of the city for finding the bandits! The mayor now trusts you and has you as his advisor. The city's population is 600 and the attack plan consists of two options and you must pick one.**

    **POSITIVE FRAMING**

1. Plan A is to openly attack the bandits and save 450 people, while not saving 150.
2. Plan B is to subtly attack with the chance of 75% that you save everyone and a 25% chance of not saving anyone.

    **NEGATIVE FRAMING**

1. Plan A kills 150 people, but doesn't kill 450.
2. Plan B has a 25% chance of killing everyone and a 75% chance of not killing anyone.